\documentclass[aps,prl,preprint,groupedaddress]{revtex4-1}

\usepackage{graphicx}
\begin{document}


\title{Percolation of Interdependent Networks with Inter-similarity}


\author{Yanqing Hu$^{1,2}$}
\author{Dong Zhou$^{3,4}$}\email{zhoudongbnu@gmail.com}
\author{Rui Zhang$^{2}$}
\author{Zhangang Han$^{3}$}
\author{Shlomo Havlin$^{4}$}
\affiliation{
  $^1$School of Mathematics, Southwest Jiaotong University, Chengdu 610031, China\\
  $^2$Levich Institute and Physics Department, City
  College of New York, New York, New York 10031, USA \\
 $^3$ Department of Systems Science, Beijing Normal University, Beijing 100875, China\\
  $^4$ Physics Department, Bar-Ilan University, Ramat Gan 52900,
  Israel }


\date{\today}

\begin{abstract}

Real data show that interdependent networks usually involve inter-similarity. Intersimilarity means that a pair of interdependent nodes have neighbors in both networks that are also interdependent (Parshani et al \cite{PAR10B}). For example, the coupled world wide port network and the global airport network are intersimilar since many pairs of linked nodes (neighboring cities), by direct flights and direct shipping lines exist in both networks. Nodes in both networks in the same city are regarded as interdependent. If two neighboring nodes in one network depend on neighboring nodes in the another we call these links common links. The fraction of common links in the system is a measure of intersimilarity. Previous simulation results suggest that intersimilarity has considerable effect on reducing the cascading failures, however, a theoretical understanding on this effect on the cascading process is currently missing. Here, we map the cascading process with inter-similarity to a percolation of networks composed of components of common links and non common links. This transforms the percolation of inter-similar system to a regular percolation on a series of subnetworks, which can be solved analytically. We apply our analysis to the case where the network of common links is an Erd\H{o}s-R\'{e}nyi (ER) network with the average degree $K$, and the two networks of non-common links are also ER networks. We show for a fully coupled pair of ER networks, that for any $K\geq0$, although the cascade is reduced with increasing $K$, the phase transition is still discontinuous. Our analysis can be generalized to any kind of interdependent random networks system.
\end{abstract}


\maketitle

\section{\label{sec1}I. Introduction}

Single isolated networks have been extensively studied in the past decade \cite{barabasi,general1, general2, general3, general4, general5, general6, general7,more1,more2,more3,more4,more5,more6,hu1}.
Recently, much interest has been devoted to interdependent networks \cite{BUL10A,PAR10A,BUL10B,SHA11,HU11,GAO10,GAO11,more7,more8,more9,more10,more11},
which can model some real world catastrophic events, such as the electrical blackout in Italy on Sep. 28th, 2003
\cite{ROSATO11} and the US-Canada Power system outage on Aug. 14th, 2003 \cite{USA-Canada1}. Failures of a small number
of power stations can cause further malfunction of nodes in their communication control network, which in turn leads to
the shutdown of power stations \cite{BUL10A,ROSATO11}. This cascading process continues until no more nodes fail due to
percolation or due to interdependence failures. In contrast to single networks where the percolation transition is continuous, in interdependent networks the transition is abrupt \cite{BUL10A,PAR10A}.

\begin{figure}
\centering
\includegraphics[scale=0.4]{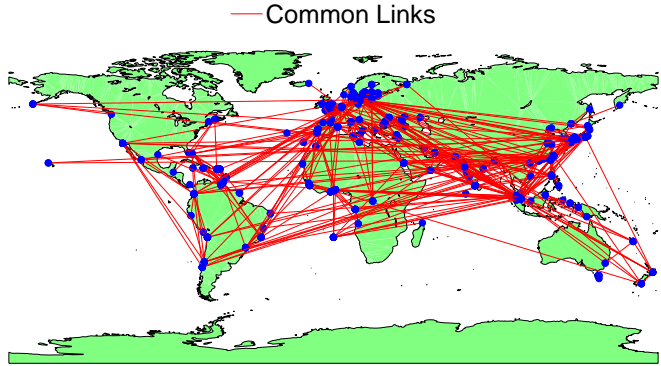}
\caption{(Color online) Common links in the interdependent geographic world wide port network and the global airport network system. In the coupled network system, we identify common links if neighboring nodes (cities with direct transportation lines) in
one network are also neighbors in the other network. For clarity, in the figure, 200 nodes are randomly chosen, and only common links attached to these nodes are plotted as solid (red) lines. Cities are shown as (blue) dots. Because two
types of nodes (ports and airports) approximately correspond to the same cities, we just use single dots to denote them in the world wide map.}
\label{fig0}
\end{figure}

Real interdependent networks are sometimes coupled according to some inter-similarity features. Intersimilarity means the tendency of neighboring nodes in one network to be interdependent of neighboring nodes in the other network. Such coupled networks,
are more robust against cascading failures than randomly coupled interdependent networks. To quantify self-similarity,
Parshani \emph{et al}. introduced the inter-clustering coefficient (ICC), which measures the average number of common
links per pair of interdependent nodes \cite{PAR10B}. Common links are defined as follows: Given two coupled networks
$A$ and $B$, and two nodes $a_k$ and $a_l$ which are linked in $A$. If their interdependent counterparts
$b_k$ (corresponds to $a_k$) and $b_l$ (corresponds to $a_l$) in $B$ are also linked (in $B$), this pair of links is called a common link.
Common links can be interpreted as follows, if two nodes in one network are linked, the tendency (probability) of their
interdependent counterparts in the other network to be linked is a measure of the inter-similarity. Thus, the density
of the common links reflects the inter-similarity of the two networks. In the extreme case where every pair of links is a common
link, the two networks are identical. In this case no cascading failure will occur, since a failure in network A will cause an identical failure in B and there will be no cascading failure feedback to A. It is therefore expected that the more common links appear in the coupled networks system, it becomes more robust. In the example of the coupled world wide port network and the airline network, illustrated
in Fig.~\ref{fig0}, the fraction of common links is 0.12 for the port network and 0.18 for the airline network \cite{PAR10B}. Therefore, developing
a method to analyze cases where certain common topologies exist in the interdependent networks can help to understand
the vulnerabilities of coupled complex systems in real world as well as for designing robust infrastructures.

In this paper, we introduce a method to analytically calculate the cascading process of
failures in interdependent networks with common links. To analyze this problem, we
consider the cluster components of the network composed of only common
links after the initial attack. We will illustrate that all nodes in such a
component will survive or fail simultaneously during the cascading process. Based on this fundamental feature, we divide the system into subnetworks according to the sizes of the components, and
then contract all nodes in each component into a single node. After contraction,
the system degenerates into two randomly coupled networks without common links, which
will be solved analytically. Here, we find the exact solution for the case where $A$
and $B$ are fully interdependent Erd\H{o}s-R\'{e}nyi (ER) networks each of average degree k and the network of
common links is also an ER network with the average degree $K$. In this case, we show
that the interdependent networks system undergoes a first order transition for all $K\geq0$.

The paper is organized as follows. In Sec. II, we introduce the model of cascading
failures in interdependent networks with common links. In Sec. III, we analyze the
process of failures and the final state where the process ends. In Sec. IV, we
derive our theoretical results and compare them with simulation results.

\section{\label{sec2}II. The Model}
\begin{figure}
\centering
\includegraphics[scale=0.6]{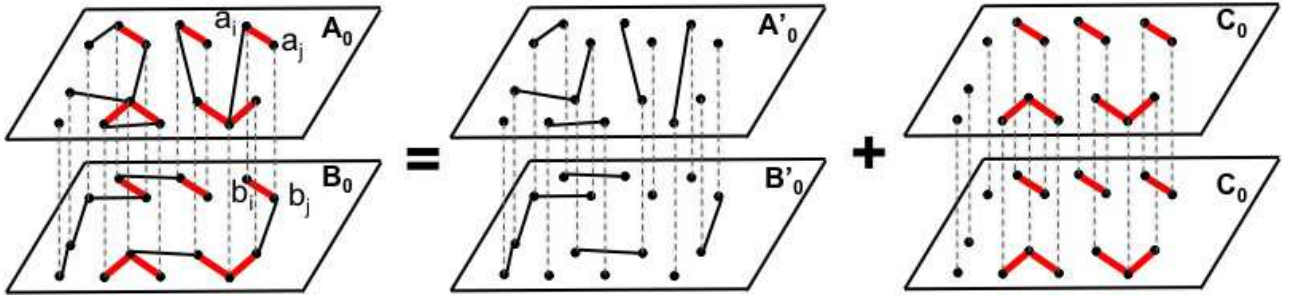}
\caption{(Color online) Decomposition of networks according to the common links.
Dots in the upper plane are the nodes from network $A_0$ and those in the lower plane are the
nodes from network $B_0$. Interdependent nodes are connected by dashed lines. Solid thick (red)
lines are the common links and other links are shown in solid thin lines. Network $C_0$ is
composed of common links. $A_0'$ and $B_0'$ are the supplementary networks with respect to
$C_0$. $a_i$ and $a_j$ are linked nodes in network $A_0$, their interdependent counterparts
$b_i$ and $b_j$ are also linked in network $B_0$. Thus the link is a common link and appears
in network $C_0$. }
\label{fig1}
\end{figure}

For simplicity and without loss of generality we analyze the percolation process in a system of two fully interdependent
networks $A$ and $B$ of the same size $N$ with no-feedback condition \cite{GAO11} in the presence of common links. The no-feedback means that, each $A$-node $a_k$ has one and only one dependency counterpart $b_k$ on
network $B$, and $b_k$ must depend only on $a_k$. Initially,
a fraction $1-p$ of $A$ nodes are removed randomly. Due to interdependency,
a corresponding fraction of $B$-nodes also fail. We denote by $A_0$ and
$B_0$ the remaining networks of size $N_{0}=pN$. Nodes of $A_0$ and $B_0$ are
represented by $a_i$ and $b_i$ respectively, $i=1,...,N_{0}$, and $a_i$ interdepends on $b_i$ for all $i=1,2,...,N$. The pair $(i,j)$ is a common link if $a_i$ is linked to $a_j$ and $b_i$ is also linked to $b_j$. We introduce a network
$C_0$ that includes all $N_{0}$ nodes but only links that are common links. This means that the nodes
$c_i$ and $c_j$ of $C_0$ will be linked if and only if both $a_i, a_j$ and $b_i, b_j$
are linked. Analogously, we define a network $C$ which is the collection of common links in the original networks $A$ and $B$. Thus network $C$ reduces to $C_0$ due to the initial attack.
As shown in Fig. ~\ref{fig1}, we denote $A_0'$ and $B_0'$ to be the networks
that composed of the same $N_{0}$ nodes in $A_0$ and $B_0$ but only those links which
are not common links. Therefore, networks $A_0$ and $B_0$ can be written as matrix
summations

\begin{equation}
 A_0=A_0'+C_0,
 B_0=B_0'+C_0.
\end{equation}

We will investigate the robustness of such a system after the initial attack. Notice that
when $C_0$ has no links since there is no common links in this system. This is the case of random coupling studied by Buldyrev \emph{et al} \cite{BUL10A}, since the probability to have a common link in random coupling approach to zero for large N. We will provide a method for analyzing
the case when network $C_0$ has a given topological structure. Let $R_{0}(m), m=1,2,\cdots,M$
be the component size distribution of $C_{0}$. That is to say, if we randomly choose
a node in network $C_{0}$, the probability that it belongs to a component of size $m$ in
network $C_{0}$ is $R_{0}(m)$. This distribution is a characterization of both the degree
and the structure of inter-similarity of the network.

The initial attack leads to failures of some other nodes in $A_{0}$ since
those nodes will lose connectivity with the giant component $A_{1}$ of
$A_{0}$ (a percolation failure). Consequently, in $B_{0}$, all nodes that
depend on those nodes that have been removed in $A_{0}$ will fail due to
interdependency relations (a dependency failure). We use $B_{1}$ to
denote the remaining nodes in $B_{0}$. Then, similarly, a percolation
failure will occur in $B_{1}$. This will induce an iterative process of
percolation failures and dependency failures in the system \cite{PAR11}.
Finally, if no further failure occurs, this cascading process will end
with a total collapse or two remaining giant components of the same size.
We are interested here in the relationship between $p$ and the size of the
final mutual giant component.

\begin{figure}
\centering
\includegraphics[scale=0.6]{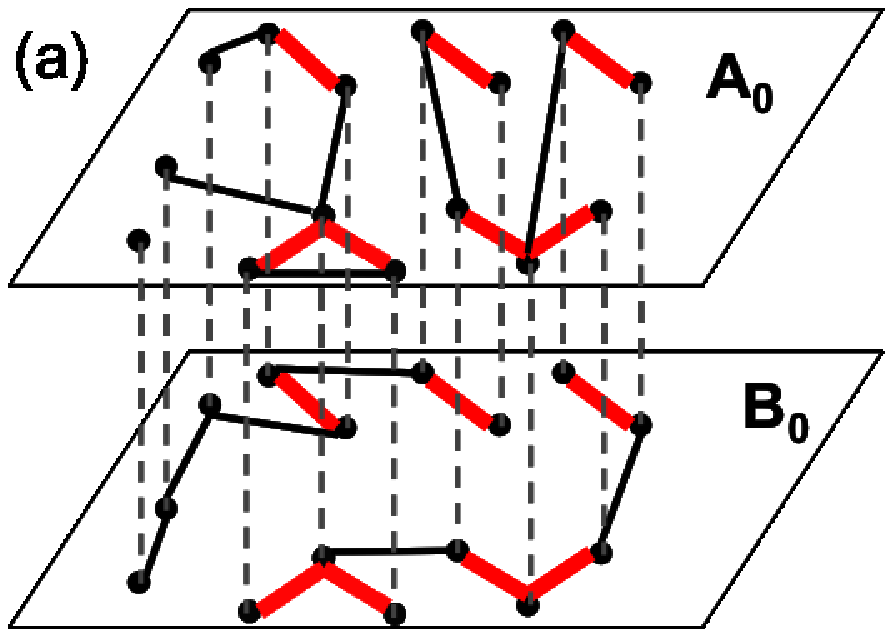} \includegraphics[scale=0.6]{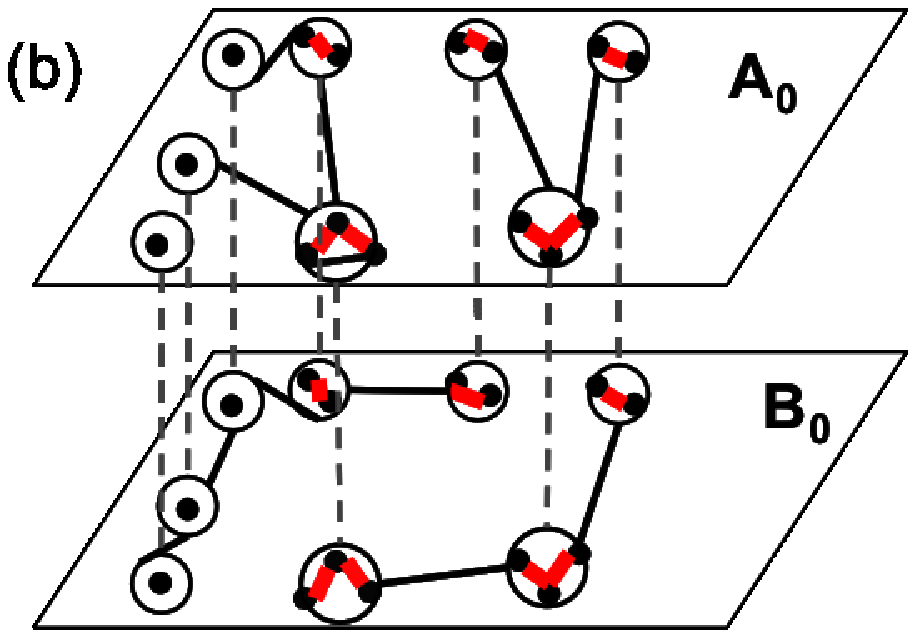} \includegraphics[scale=0.6]{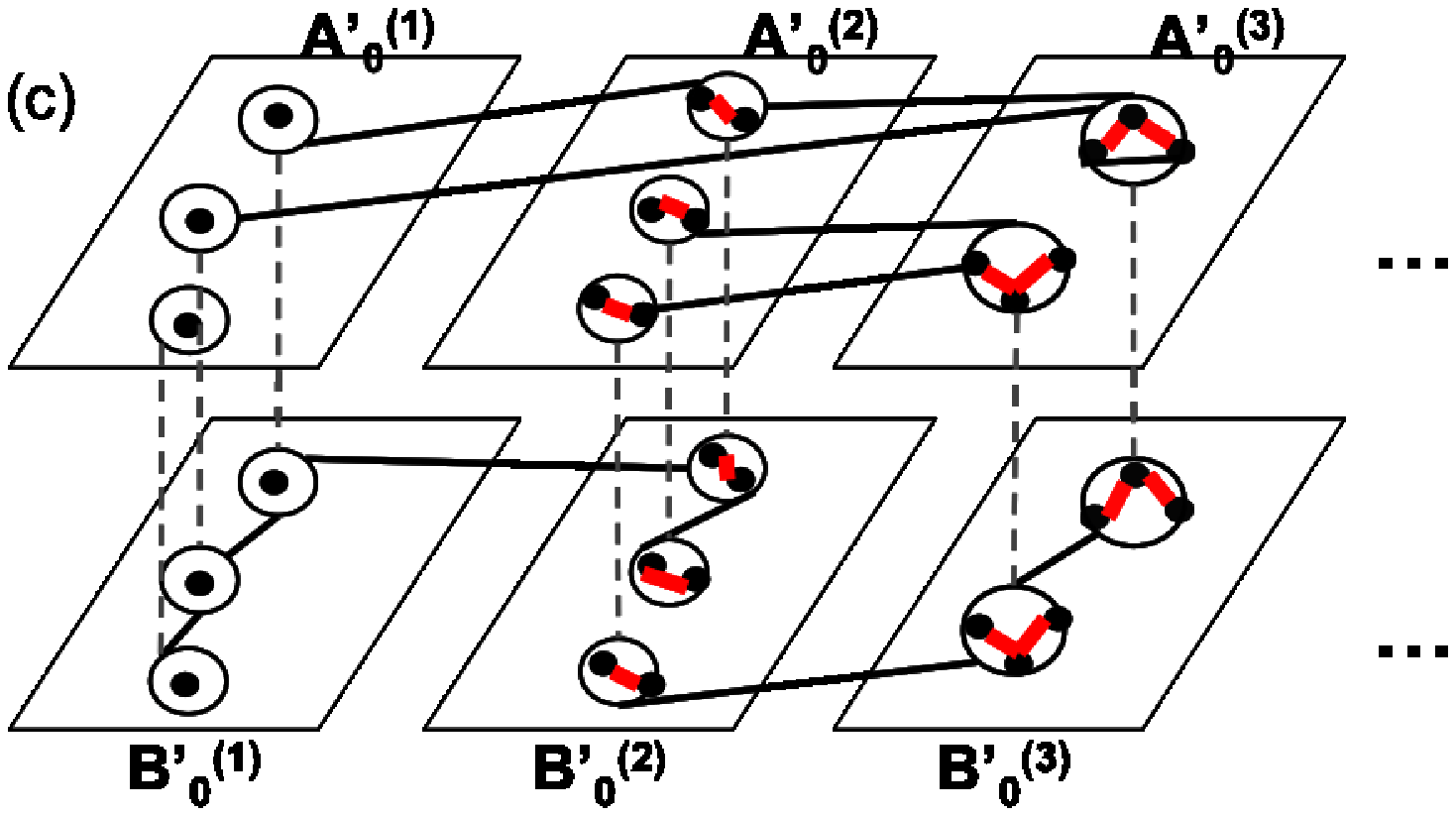}
\caption{(Color online) A sketch of the method of dividing and
contracting the system after the initial attack. (a) The remaining interdependent networks $A_{0}$
and $B_{0}$ right after the initial attack. Thick (red) lines are the common links. Other links are shown in (black) solid thin lines. Interdependent nodes are connected by dashed lines.
(b) Nodes in the same component of $C_0$ are contracted. Nodes that are within one circle belong to the same component in network $C_0$. Because of the first principle we propose, all the nodes from one component either survive together or die together. Thus the component can be regarded as a super node (if survived). The links between nodes from different components are now links between the two components.
(c) The contracted network is further decomposed into subnetworks according to the component size in $C_0$. For example, $A'^{(2)}_0$ is the collection of super nodes of component size 2 in $A_{0}$, etc.}
\label{fig2}
\end{figure}

Notice that during the cascading process, if a node in a component of $C_0$
survives, the whole component will survive, and if a node in a component
fails, the whole component will fail.
Inspired by this basic fact, as shown in Fig. \ref{fig2}, we divide networks $A_{0}$
and $B_{0}$ respectively according to the component sizes in network $C_{0}$.
That is to say, in network $A_{0}$, all those nodes that belong to components
of size $m=1,2,\cdots,M$ in $C_{0}$ compose a subnetwork denoted by $A_{0}^{(m)}$.
Here, $M$ is the largest component size of network $C_{0}$. Network $B_{0}$
can be divided into $B_{0}^{(m)}$, analogously. In this way, the size of each subnetwork $A_{0}^{(m)}$ or $B_{0}^{(m)}$ is $N_{0}^{(m)}=R_{0}(m)N_{0}$, where $m=1,2,\cdots,M$.

As depicted in Fig. ~\ref{fig2}, we also contract $A_{0}$ and $B_{0}$ according to the components of
network $C_{0}$. In other words, in network $A_{0}$, all $m$-nodes that
belong to each component of size $m$ in $C_{0}$ are merged into a single
new node, and all links connected to at least one of these $m$ nodes are
also merged on the new node. All of these merged nodes and links form a
contracted network $A'_{0}$ with subnetworks $A'^{(m)}_{0}, m=1,2,\cdots,M$.
Network $B_{0}$ is contracted similarly. In this way, the size of each
subnetwork $A'^{(m)}_{0}$ or $B'^{(m)}_{0}$ is just the number of components
of size $m$ in network $C_{0}$, that is $N'^{(m)}_{0}=\frac{R_{0}(m)N_{0}}{m}, m=1,2,\cdots,M$.
And the size of $A'_{0}$ or $B'_{0}$ is
$N'_{0}=\sum\limits_{m=1}^{M}N^{(m)}_{0}=\sum\limits_{m=1}^{M}\frac{R_{0}(m)N_{0}}{m}=N_{0}\cdot\sum\limits_{m=1}^{M}\frac{R_{0}(m)}{m}$.
We denote $\langle m\rangle=\left(\sum\limits_{m=1}^{M}\frac{R_{0}(m)}{m}\right)^{-1}$
as the average component size in $C_{0}$. Thus, we have $N'_{0}=\frac{N_{0}}{\langle m\rangle}$.

As shown in Fig.~\ref{fig2}(c), common links do not exist any more in the
contracted system, because each common link always lies inside a component
of network $C_{0}$. In fact, after the initial attack, the cascading process
in the contracted system is equivalent to the cascading on the original system.
Therefore, we only need to focus on the cascade process in the contracted system.

\section{III. Theoretical Approach}
Here, we exhibit, step-by-step, the theoretical analysis for the cascading process
starting from $A'_{0}$ and $B'_{0}$. In the first stage, the size of the remaining
functional giant component $A'_{1}$ can be obtained using the method proposed
by Leicht and D'Souza \cite{LEI09}. We regard $A'_{0}$ as a system of $M$ coupled subnetworks
$A'^{(m)}_{0}$, $m=1,2,\cdots,M$. If the degree distribution $p^{mm'}_{A'_{0}}$
($mm'$-degrees) from a randomly chosen node in $A'^{(m)}_{0}$ to all nodes in
$A'^{(m')}_{0}$ can be exactly evaluated, then the whole system can be described using
multi-variable generating functions. Usually, we analyze the cascading process by three steps to obtain the
recursive system \cite{BUL10A} .

We use $g^{(m)}_{A'_{0}}(p_{1},p_{2},\cdots,p_{M})$ to denote the fraction of
nodes in the giant component of $A'^{(m)}_{0}$ after randomly removing a fraction
$1-p_{m'}$ of nodes in each subnetwork $A'^{(m')}_{0}$, $m'=1,2,\cdots,M$. Here, we contract the two coupled networks
after the initial attacking. It implies that $p_{m'}=0$ at the beginning of the cascading process on the contracted two coupled networks.
Thus, the remaining functional part in each subnetwork is $\psi^{(m)}_{1}=g^{(m)}_{A'_{0}}(1,1,\cdots,1)$
in the first stage.

The second stage \cite{BUL10A} is equivalent to randomly attacking a fraction $1-\psi^{(m)}_{1}$
of nodes in each subnetwork $B'^{(m)}_{0}$, $m=1,2,\cdots,M$. We let
$\phi'^{(m)}_{1}=\psi^{(m)}_{1}$. Therefore, the remaining giant component of
$B'_{0}$ is $\phi^{(m)}_{1}=\phi'^{(m)}_{1}\cdot g^{(m)}_{B'_{0}}(\phi'^{(1)}_{1},\phi'^{(2)}_{1},\cdots,\phi'^{(M)}_{1})$.

The third stage is equivalent to randomly removing a fraction
$1-g^{(m)}_{B'_{0}}(\phi^{(1)}_{1},\phi^{(2)}_{1},\cdots,\phi^{(M)}_{1})$ of
nodes in $A'^{(m)}_{0}$. We let $\psi'^{(m)}_{2}=g^{(m)}_{B'_{0}}\cdot (\phi^{(1)}_{1},\phi^{(2)}_{1},\cdots,\phi^{(M)}_{1})$,
$m=1,2,\cdots,M$. Thus, the remaining fraction in $A'_{0}$ is
$\psi^{(m)}_{2}=\psi'^{(m)}_{2}\cdot g^{(m)}_{A'_{0}}(\psi'^{(1)}_{2},\psi'^{(2)}_{2},\cdots,\psi'^{(M)}_{2})$.

Generally, we have
$\psi'^{(m)}_{n}=g^{(m)}_{B'_{0}}\cdot (\phi'^{(1)}_{n-1},\phi'^{(2)}_{n-1},\cdots,\phi'^{(M)}_{n-1})$, and
$\psi^{(m)}_{n}=\psi'^{(m)}_{n}\cdot g^{(m)}_{A'_{0}}(\psi'^{(1)}_{n},\psi'^{(2)}_{n},\cdots,\psi'^{(M)}_{n})$;
$\phi'^{(m)}_{n}=g^{(m)}_{A'_{0}}(\psi'^{(1)}_{n},\psi'^{(2)}_{n},\cdots,\psi'^{(M)}_{n})$, and
$\phi^{(m)}_{n}=\phi'^{(m)}_{n}\cdot g^{(m)}_{B'_{0}}(\phi'^{(1)}_{n},\phi'^{(2)}_{n},\cdots,\phi'^{(M)}_{n})$,
$m=1,2,\cdots,M$.

In the final stage, where the process of cascading failures ceases,
we have $\psi'^{(m)}_{n}=\psi'^{(m)}_{n-1}=\phi'^{(m)}_{n}=\phi'^{(m)}_{n-1}$ for
all $m$. Let $x_{m}=\psi'^{(m)}_{n}$, and $y_{m}=\phi'^{(m)}_{n}$. We
arrive a system of $x_{m}$ and $y_{m}$:

\begin{equation}
x_{m}=g^{(m)}_{A'_{0}}(y_{1},y_{2},\cdots,y_{M}),
y_{m}=g^{(m)}_{B'_{0}}(x_{1},x_{2},\cdots,x_{M}),
\end{equation}

\noindent $m=1,2,\cdots,M$.

\section{\label{sec4}IV. Analytical Solution}

This system can be analytically solved using $M$-variant generating
functions. Similar to Ref. \cite{LEI09}, for a system of $M$ interconnected
subnetworks $A'^{(m)}_{0}$, we define the generating function for the
degree distributions for each subnetwork as

\begin{equation}
G^{(m)}_{A'_{0}}(\xi_{1},\xi_{2},\cdots,\xi_{M})=\sum\limits_{k_{1},k_{2},\cdots,k_{M}}p^{(m)}_{k_{1},k_{2},\cdots,k_{M}}\cdot\xi^{k_{1}}_{1}\xi^{k_{2}}_{2}\cdots\xi^{k_{M}}_{M},
\end{equation}

\noindent where $p^{(m)}_{k_{1},k_{2},\cdots,k_{M}}$ is the probability that a
randomly chosen node in $A'^{(m)}_{0}$ has $k_{m'}$ $mm'$-degrees.
Moreover, the generating function for the underlying branching processes for
each subnetwork is

\begin{equation}
G^{(mm')}_{A'_{0}}(\xi_{1},\xi_{2},\cdots,\xi_{M})=\frac{\frac{\partial}{\partial\xi_{m'}}G^{(m)}_{A_{0}}(\xi_{1},\xi_{2},\cdots,\xi_{M})}{\frac{\partial}{\partial\xi_{m'}}G^{(m)}_{A_{0}}(\xi_{1},\xi_{2},\cdots,\xi_{M})\left.\right|_{\xi_{1}=\xi_{2}=\cdots=\xi_{M}=1}},
\end{equation}

\noindent where $m'=1,2,\cdots,M$. Then, the fraction of nodes in the giant component
after randomly removing a fraction $1-p_{m}$ of nodes in each subnetwork $A'^{(m)}_{0}$ is

\begin{equation}
g^{(m)}_{A'_{0}}(p_{1},p_{2},\cdots,p_{M})=1-G^{(m)}_{A'_{0}}(1-p_{1}\cdot(1-u_{1m}),1-p_{2}\cdot(1-u_{2m}),\cdots,1-p_{M}\cdot(1-u_{Mm})).
\end{equation}

\noindent Here, $u_{m'm}$ satisfies:

\begin{equation}
u_{m'm}=G^{m'm}_{A'_{0}}(1-p_{1}\cdot(1-u_{1m'}),1-p_{2}\cdot(1-u_{2m'}),\cdots,1-p_{M}\cdot(1-u_{Mm'})),
\end{equation}
where, $m,m'=1,2,\cdots,M.$
For network $B'_{0}$, we can define the analogous generating functions and obtain similarly
the giant component size.

For simplicity, we assume that all $mm'$-degree distributions in $A'_{0}$ and $B'_{0}$
are Poisson distributions, whose average degrees are $k^{A'_{0}}_{mm'}$ and
$k^{B'_{0}}_{mm'}$, respectively. For example, if both $\tilde{A}$ and $\tilde{B}$ are
Erd\H{o}s-R\'{e}nyi (ER) networks with average degrees $a$ and $b$,
respectively, and the initial attack on network $A$ is random, then these
$mm'$-degrees in $A_{0}$ have Poisson distributions for all $m$ and $m'$.
Then, according to the result in Ref. \cite{LEI09}, we have

\begin{equation}
G^{m}_{A'_{0}}=G^{mm'}_{A'_{0}}=\exp\left(-\sum\limits_{m'=1}^{M}k^{A'_{0}}_{mm'}(1-\xi_{m'})\right),
G^{m}_{B'_{0}}=G^{mm'}_{B'_{0}}=\exp\left(-\sum\limits_{m'=1}^{M}k^{B'_{0}}_{mm'}(1-\xi_{m'})\right).
\end{equation}

\noindent Here, the average $mm'$-degrees in $A'_{0}$ and $B'_{0}$ are $k^{A'_{0}}_{mm'}=amp\cdot R_{0}(m')$
and $k^{B'_{0}}_{mm'}=bmp\cdot R_{0}(m')$, respectively. Notice that, in this case,
$u_{m1}=u_{m2}=\cdots=u_{mM}\triangleq u_{m}$, $m=1,2,\cdots,M$. Therefore,

\begin{equation}
g^{(m)}_{A'_{0}}(p_{1},p_{2},\cdots,p_{M})=1-u_{m}(p_{1},p_{2},\cdots,p_{M}),
\end{equation}

\noindent where $u_{m}(p_{1},p_{2},\cdots,p_{M})$, $m=1,2,\cdots,M$ is the
solution of the following set of equations:

\begin{equation}
u_{m}=G^{m}_{A'_{0}}(1-p_{1}\cdot(1-u_{1}),1-p_{2}\cdot(1-u_{2}),\cdots,1-p_{M}\cdot(1-u_{M}))=\exp\left(-\sum\limits_{m'=1}^{M}k^{A'_{0}}_{mm'}\cdot p_{m'}(1-u_{m'})\right),
\end{equation}
where, $ m=1,2,\cdots,M.$
\noindent Similarly, for network $B'_{0}$,

\begin{equation}
g^{(m)}_{B'_{0}}(p_{1},p_{2},\cdots,p_{M})=1-v_{m}(p_{1},p_{2},\cdots,p_{M}),
\end{equation}

\noindent where $v_{m}$ satisfies:

\begin{equation}
v_{m}=G^{m}_{B'_{0}}(1-p_{1}\cdot(1-v_{1}),1-p_{2}\cdot(1-v_{2}),\cdots,1-p_{M}\cdot(1-v_{M}))=\exp\left(-\sum\limits_{m'=1}^{M}k^{B'_{0}}_{mm'}\cdot p_{m'}(1-v_{m'})\right)
\end{equation}
where, $ m=1,2,\cdots,M$. The system of the final stage can be written as
\begin{equation}
x_{m}=1-u_{m}(y_{1},y_{2},\cdots,y_{M}),
y_{m}=1-v_{m}(x_{1},x_{2},\cdots,x_{M}).
\end{equation}

\noindent By excluding $x_{m}$ and $y_{m}$, we finally obtain

\begin{equation}
u_{m}=\exp\left(-\sum\limits_{m'=1}^{M}k^{A'_{0}}_{mm'}(1-u_{m'})(1-v_{m'})\right),
v_{m}=\exp\left(-\sum\limits_{m'=1}^{M}k^{B'_{0}}_{mm'}(1-u_{m'})(1-v_{m'})\right).
\end{equation}

\noindent Therefore,

\begin{equation}
u_{m}=\exp\left(-amp\sum\limits_{m'=1}^{M}R_{0}(m')(1-u_{m'})(1-v_{m'})\right),
v_{m}=\exp\left(-bmp\sum\limits_{m'=1}^{M}R_{0}(m')(1-u_{m'})(1-v_{m'})\right),
\label{eq14}
\end{equation}

\noindent where $m=1,2,\cdots,M$.

By solving this system, we can get $\mu^{(m)}_{\infty}=(1-u_{m})(1-v_{m})$,
$m=1,2,\cdots\,M$. This is the fraction of the mutual giant component
in each subnetwork $A'^{(m)}_{0}$ or $B'^{(m)}_{0}$. The fraction of
the mutual giant component in the original system of $A$ and $B$ is

\begin{equation}
\mu_{\infty}=\frac{\sum\limits_{m=1}^{M}\mu^{(m)}_{\infty}\cdot N'^{(m)}_{0}\cdot m}{N}=p\cdot\sum\limits_{m=1}^{M}\mu^{(m)}_{\infty}R_{0}(m).
\end{equation}

Notice that in Eq.~(\ref{eq14}), $u_{m}=u^{m}_{1}$, and $v_{m}=u^{mb/a}_{1}$,
$m=1,2,\cdots,M$. Therefore, the system can be simplified to a single equation
for $u_{1}$,

\begin{equation}
u_{1}=\exp\left(-ap\sum\limits_{m'=1}^{M}R_{0}(m')(1-u^{m'}_{1})(1-u^{m'b/a}_{1})\right).
\label{eq16}
\end{equation}

\noindent Thus, the fraction of the mutual giant component becomes

\begin{equation}\label{eq17}
\mu_{\infty}=p\cdot\sum\limits_{m=1}^{M}R_{0}(m)(1-u^{m}_{1})(1-u^{mb/a}_{1})=\frac{-\log u_{1}}{a}.
\end{equation}

\section{\label{sec5}V. Results}

One trivial solution of Eq.~(\ref{eq16}) is $u_{1}=1$. In some cases, other nontrivial
solutions exist in the interval $[0,1)$. The smallest solution $u_{1min}$ corresponds
to the size of the final mutual giant component ${-\log u_{1min}}/{a}$. Consider
$F_{2}(u_{1})=\exp\left(-ap\sum\limits_{m'=1}^{M}R_{0}(m')(1-u^{m'}_{1})(1-u^{m'b/a}_{1})\right)$ and $F_{1}(u_{1})=u_{1}$.
Then the critical point $u_{1c}$ and $p^{I}_{c}$ is where a nontrivial solution that
satisfies: $F_{1}(u_{1c})=F_{2}(u_{1c})$ and $F'_{1}(u_{1c})=F'_{2}(u_{1c})$ emerges.
Note that, all the analysis is done here on the contracted network system after the
initial random removal, which means there is no initial attacking on the contracted system.
If $M$ is finite, at the solution $u_{1}=1$, we have $F_{1}(1)=F_{2}(1)=1$, but $F'_{1}(1)=1\neq F'_{2}(1)=0$.
This means these two curves cannot be tangent to each other at $u_{1}=1$. Therefore, $u_{1}=1$ cannot be a critical value for a second order transition, and only first order phase transitions at $u_{1c}<1$ occur in systems with a finite $M$.

\begin{figure}
\includegraphics[width=0.60\textwidth]{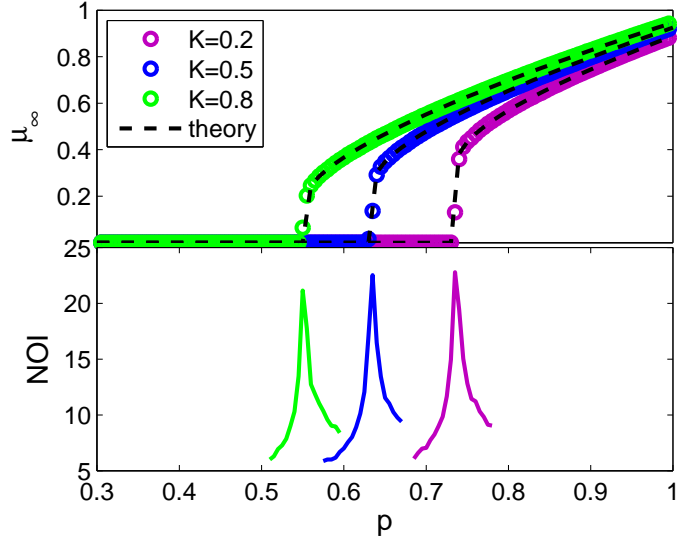}
\caption{(Color online) Comparison of the simulation results (symbols) for $\mu_{inf}$ (top figure) against the theoretical results (curves)
when network $C_{0}$ is an ER network. Here, we choose $a=b=3$, $K=0.2, 0.5, 0.8$ and $N=100,000$. We also show (bottom figure) the number of iterative failures (NOI) in the simulation for the same values of $K$. Usually, the largest peak of NOI values indicates the critical point at the first order phase transition \cite{PAR11}. We can see the excellent agreement between the theoretical and simulation results.}
\label{fig3}
\end{figure}

\begin{figure}
\includegraphics[width=0.60\textwidth]{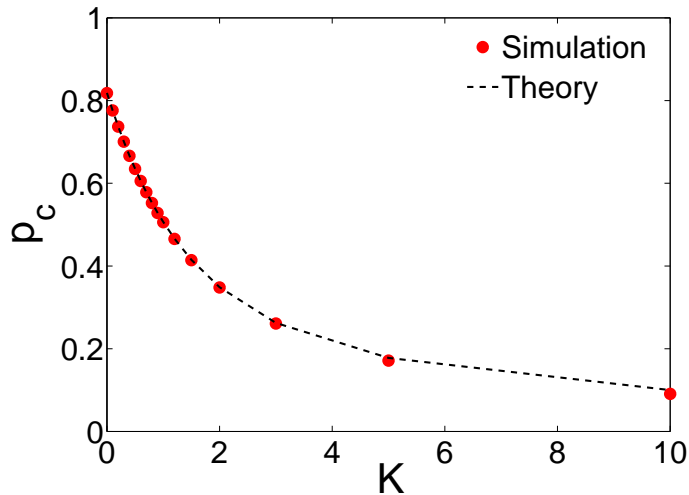}
\caption{The dependence of $p_{c}$ on average degree $K$ of common links for interdependent ER networks systems.
The average degrees of networks $A_0$ and $B_0$ are $a=b=3$, and that of $C_0$ is $K$. The network size is $N=300,000$, and the number of realizations in simulations is 100. The curve is the theoretical results (using $M_{max}=50$), and (red) full circles are the corresponding simulation results.}
\label{fig4}
\end{figure}

\begin{figure}
\includegraphics[width=0.60\textwidth]{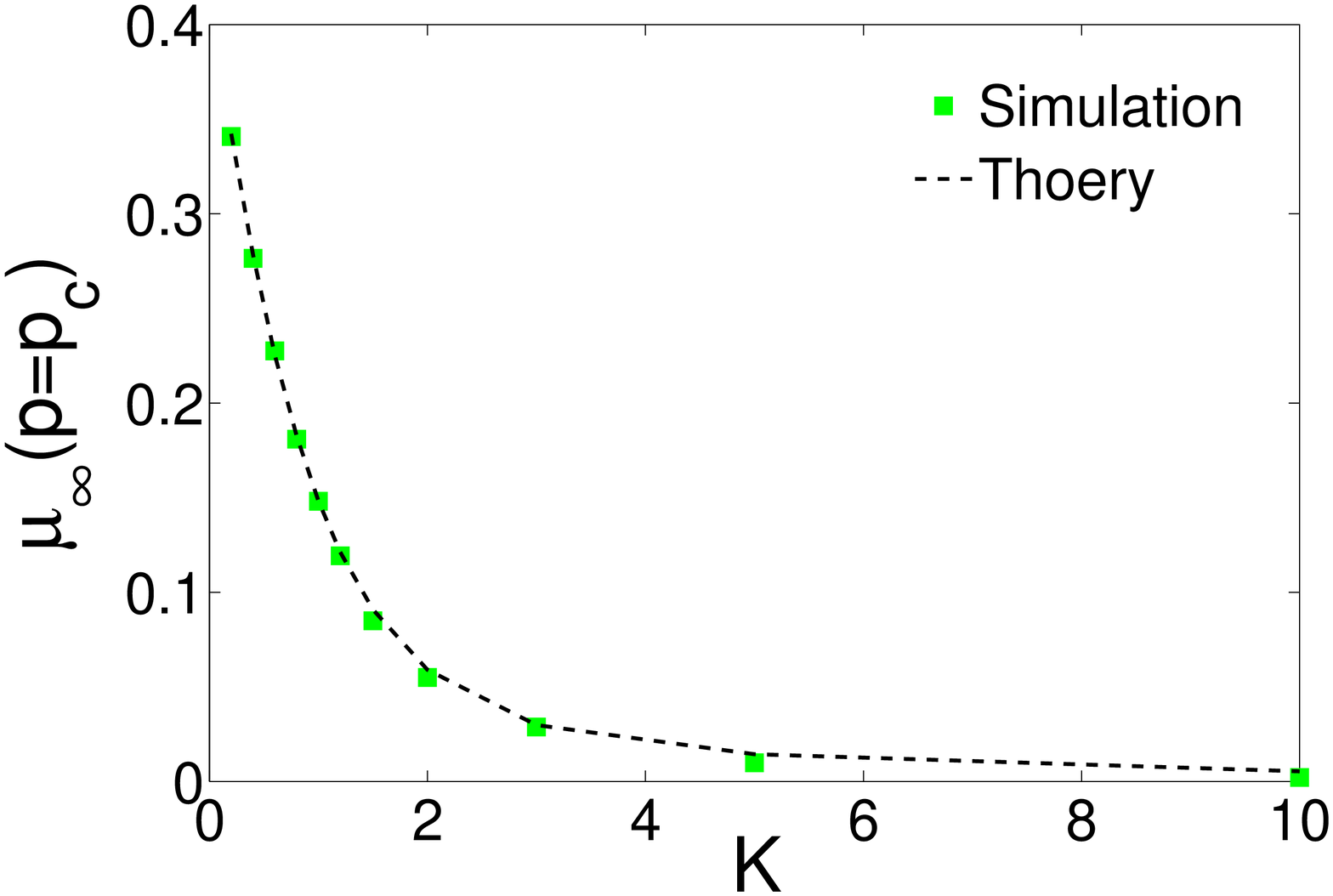}
\caption{The dependence of the height of the jump at the critical point of the first order phase transitions on the average degree $K$ of network $C_0$. The average degrees of networks $A_0$ and $B_0$ are $a=b=3$. The network size is $N=100,000$, and the number of realizations in simulations is 100. The dashed curve is the theoretical result ($M_{max}=50$), and (green) solid squares are the corresponding simulation results at the critical points. From the shape of the curve, we can see that the height of the jump at the critical point does not reach 0 for a very large $K$, which strongly supports that if the two mutually depending networks are not identical, the phase transition is always discontinuous.}
\label{fig5}
\end{figure}

Here, we further investigate the case where $C$ (network composed of common links) is an ER network with an average degree $K$.
After the random initial attack, $C_{0}$ is also an ER network, whose average degree
becomes $Kp$. The component size distribution of $C_{0}$ is $R_{0}(m)=\frac{1}{m!}\cdot(mKp)^{m-1}\cdot e^{-mKp}, m=1,2,\cdots$.
In this case, $M$ should be in principle infinite. However, when we substitute this distribution into
Eq.~(\ref{eq16}), it is appropriate to use a truncation $m=1,2,\cdots,M<\infty$ on the
infinite sum, since $R_{0}(m)$ decays exponentially, and the largest cluster of network $C_{0}$
cannot be as large as $O(N)$ when $Kp<1$. This means $p_{c}<1/K$.
Therefore, no matter for $K<1$ or $K\geq 1$, we can use Eq.~(\ref{eq16}) directly to determine $p_{c}$.
The theoretical results are in excellent agreement with simulations and are shown in Fig. ~\ref{fig3},
Fig. ~\ref{fig4} and Fig. ~\ref{fig5}. Also, notice that for $a=b=0$ and $K\geq 1$, the system will
behave like a single ER network and will have a second order transition. Surprisingly, our result indicates
that for any $a, b>0$, which means the two networks are not identical, the equation describing the system will become Eq. (\ref{eq16}),
and the transition will become suddenly from second order when $a=b=0$ to first order.

\section{VI. Summary}

In this paper we provide an exact solution for interdependent networks with common links (representing inter-similarity in the system),
which can be found in many real world network systems. We treat the components composed of inter-
similar links as a new kind of nodes, and these new nodes form a new mutually interdependent network
system with degree correlation, which comes from the correlation between component sizes. In order
to deal with this kind of degree correlation, we decompose the new network system into a series of
subnetworks according to their component sizes. That is, the new node correspond to the same component
size in each of subnetworks respectively. Then we employ a high dimensional generating function to
describe this system and obtain the exact percolation equations which can be solved numerically. If
the two mutually interdependent networks are fully inter-similar or identical ($a=b=0$), we know
that the percolation is exactly the same with that on a single network and must be a second order
phase transition. From the above analysis, we surprisingly find that when the two mutually
interdependent networks are not identical ($a, b>0$), the transition is totally different from single networks
and is always of first order.

\section{\label{sec7}VII. Acknowledgements}
 We thanks Prof. Yiming Ding and Dr. Jianxi Gao for useful discussions. This work is mainly supported by the NSFC Grants No. 61203156. DZ thanks the NSFC Grants No. 61074116 and the Fundamental Research Funds for the Central Universities of China. SH thanks DTRA, the LINC EU projects and EU-FET project MULTIPLEX No. 317532, the DFG and the Israel Science Foundation for
support.

\bibliography{manuscript1}

\end{document}